\documentclass[prd, amsfonts, twocolumn, nofootinbib, showpacs]{revtex4}
\usepackage{graphicx}
\usepackage{epsfig}
\usepackage{color}
\usepackage{amsmath}
\usepackage{amssymb}
\newcommand{\be}{\begin{equation}}
\newcommand{\ee}{\end{equation}}
\newcommand{\bea}{\begin{eqnarray}}
\newcommand{\eea}{\end{eqnarray}}

\newcommand{\gapp}{\mathrel{\raise.3ex\hbox{$>$}\mkern-14mu
\lower0.6ex\hbox{$\sim$}}}
\newcommand{\lapp}{\mathrel{\raise.3ex\hbox{$<$}\mkern-14mu
\lower0.6ex\hbox{$\sim$}}}
\def\bbox{{\,\lower0.9pt\vbox{\hrule \hbox{\vrule height 0.2 cm
\hskip 0.2 cm \vrule  height 0.2 cm}\hrule}\,}}

\begin{document}
\title{A new wormhole solution in de Sitter space}
\author{De-Chang Dai$^{1,4}$\footnote{communicating author: De-Chang Dai,\\ email: diedachung@gmail.com\label{fnlabel}}, Djordje Minic$^2$, Dejan Stojkovic$^3$, }
\affiliation{$^1$ Center for Gravity and Cosmology, School of Physics Science and Technology, Yangzhou University, 180 Siwangting Road, Yangzhou City, Jiangsu Province, P.R. China 225002 }
\affiliation{ $^2$ Department of Physics, Virginia Tech, Blacksburg, VA 24061, U.S.A. }
\affiliation{ $^3$ HEPCOS, Department of Physics, SUNY at Buffalo, Buffalo, NY 14260-1500, U.S.A.}
\affiliation{ $^4$ CERCA/Department of Physics/ISO, Case Western Reserve University, Cleveland OH 44106-7079}

\begin{abstract}
\widetext

We present a new wormhole solution connecting two points of the same universe separated by finite distance. Virtually all the existing solutions connect two disconnected universes, or two points of the same universe that are infinitely far away. 
We construct our solution by placing two black holes at the antipodes
of the closed de Sitter space with a matter shell between them. The
gravitational action of the matter shell and cosmological constant
counteracts
attractive gravity between the black holes and makes the whole configuration static. The whole space outside the wormhole mouths is causally connected, even though the wormhole is not traversable. The stress energy tensor corresponds to de Sitter vacuum everywhere outside of the black holes except at the equator where we match the black hole spacetimes, where a shell with positive energy density appears.
We discuss the physical relevance of this solution in various contexts, including the cosmological constant problem.
\end{abstract}

\maketitle


The study of wormholes has a long and distinguished history. The original wormhole solution was discovered by
Einstein and Rosen (ER) in 1935 \cite{ER}. In the 1950s and 1960s John Wheeler \cite{geons}
and collaborators have emphasized the importance of wormholes (and topology change) in quantum gravity, as insightfully reviewed in \cite{wheeler}. Then in the 1980s two parallel developments stressed the role of wormholes in fundamental physics: in the first development, Baum \cite{Baum:1984mc}, Hawking \cite{Hawking:1984hk} and Coleman \cite{Coleman:1988tj} focused on the role of topology change in Euclidean quantum gravity (see \cite{gibbons} for a review),
and they speculated that this process is crucial for the possible fix of fundamental constants in nature, and in particular,
the cosmological constant. Around the same time Kip Thorne and collaborators realized that it was possible to construct
``traversable'' wormhole solutions \cite{Morris:1988cz, Morris:1988tu}. (For an in-depth review of this latter work
consult \cite{visser}.) More recently there has been a lot
of activity on the subject of wormholes and quantum entanglement (in the form of the Einstein-Podolski-Rosen (EPR) set-up) since the ER=EPR proposal \cite{Maldacena:2013xja} (see also, \cite{holland}).

Such possible connections between topology change in quantum gravity
and the (distribution of) values of fundamental constants, as
well as the conjectured relation between wormholes and quantum entanglement represent the main motivation for our present work.
In this letter we present a new exact solution to vacuum Einstein's equations describing a wormhole connecting two causally connected points of the same universe separated by finite distance.  
This new solution is obtained by placing two black holes at the antipodes
of the closed de Sitter space with a matter shell between them. In this situation the
gravitational action of the matter shell and cosmological constant
counteracts
attractive gravity between the black holes and makes the whole configuration static. An interesting feature of this solution (and what makes it substantially different from the maximal extension of Schwarzschild de Sitter black hole) is that causal communication is in principle possible across the equator since the cosmological de Sitter horizon does not have to be crossed. We show that the metric is non-singular at the equator, but a shell with positive energy density appears there.
Motivated by this solution, we then discuss its physical relevance in the
contexts of the dS/CFT \cite{Hull:1998vg, Strominger:2001pn, Balasubramanian:2001rb, Witten:2001kn, Balasubramanian:2001nb, Balasubramanian:2002zh} and AdS/CFT \cite{Maldacena:1997re, Gubser:1998bc, Witten:1998qj} duality and the possible relationship of wormhole configurations and quantum entanglement, especially in the setting of the Baum-Hawking-Coleman proposal.

\begin{figure}
   \centering
\includegraphics[width=7cm]{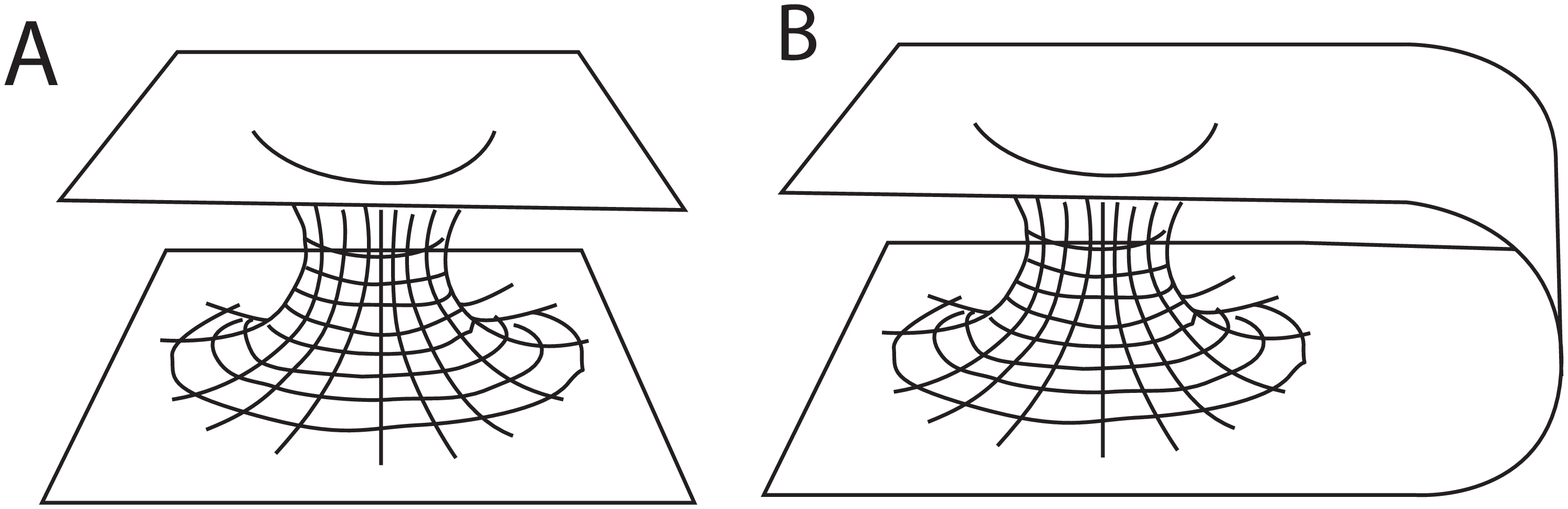}
\caption{A is the original wormhole solution constructed by Einstein and Rosen. The wormhole connects two different universes. B is the same solution as A, but the universes are connected at infinity. While the whole construct looks as if it were one single universe, the black holes are still infinitely far apart.
}
\label{wormhole}
\end{figure}

The first wormhole solution was originally constructed by Einstein and Rosen in \cite{ER}.
This wormhole connects two spacetime points through two black holes. If we start from the static black hole metric in the Schwarzschild form
\begin{equation}
ds^2=-(1-\frac{2M}{r})dt^2+\frac{dr^2}{1-\frac{2M}{r}}+r^2d\Omega
\end{equation}
and apply a simple coordinate transformation, $u^2=r-2M$, we find
\begin{equation}
ds^2=-\frac{u^2}{u^2+2M}dt^2+4(u^2+2M)du^2+(u^2+2M)^2d\Omega .
\end{equation}
This metric contains two asymptotically flat spacetimes, $u>0$ and $u<0$, which are connected at $u=0$. The geometry is shown in Fig.~\ref{wormhole} A. In this representation, the wormhole connects two different universes, and as such it does not allow for shortcuts connecting separate points in the same universe. To circumvent this feature, these two universes are sometimes artificially connected at infinity, to make the whole construct look as if it were one single universe, for example as in
Fig.~\ref{wormhole} B. Since the geometry is static, one can argue that two distant points are in causal contact since the signal has infinite time to travel between them (that is, outside the wormhole).  However, since the two black holes are still infinitely far away, they cannot have any useful communication or interaction outside the wormhole. This might not be a serious problem; however it renders this solution useless when such interaction is needed, for example, as in the ER=EPR conjecture \cite{Maldacena:2013xja}.

\begin{figure}
   \centering
\includegraphics[width=7cm]{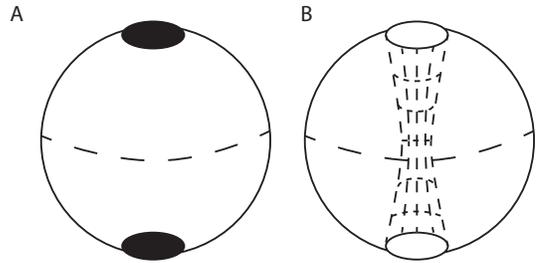}
\caption{A represents closed de Sitter space with two black holes located at the north pole and south pole respectively. B represents the same black holes connected by a wormhole.  There is a mass shell at the equator marked with a dashed line. Now the causal signal can go either through the wormhole or across the equator which connects the north and south hemispheres. 
}
\label{wormhole3}
\end{figure}

In the present paper, we want to construct a wormhole solution which connects two black holes that are finite distance apart, and which can still communicate outside the wormhole.
At first glance, it appears that we need a time dependent solution, since two black holes that are finite distance apart are always attracted to each other. One can achieve a static configuration by assigning some charge to the black holes to counteract gravity. Such a solution might exist, but it is not clear if it can be found in analytic form. Alternatively, to overcome this problem, we consider a static closed universe. We place two Schwarzschild black holes at the two antipodes (say the north and south poles, respectively), as in Fig.~\ref{wormhole3} A. The black holes still gravitationally attract each other and make a static solution impossible to find. Therefore, we work in de Sitter space endowed with positive vacuum energy density (cosmological constant), which produces a repulsive force that could balance the gravitational attraction of two black holes.
Also, we introduce a matter shell between the black holes.
Thus, the new solution is obtained by placing two black holes at the antipodes
of the closed de Sitter space with a matter shell between them. In this situation the
gravitational action of the matter shell and cosmological constant
counteracts
attractive gravity between the black holes and makes the whole configuration static. 

For our purpose, we write a metric for the closed spherically symmetric de Sitter space in the form
\begin{equation}
ds^2=-A(\lambda)dt^2+B(\lambda)d\lambda^2+r^2(\lambda) d\Omega ,
\label{full-metric}
\end{equation}
where $r(\lambda)$ is the radial coordinate defined as $\sqrt{\frac{S}{4\pi}}$, while $S$ is the surface of a two-sphere with the center located at the north pole. The Einstein tensor for this metric is

\begin{eqnarray} \label{et}
G^t_t&=&\frac{2r''Br+Br'{}^2-rr'B'-B^2}{r^2B^2}\\
G^\lambda_\lambda &=&\frac{Ar'{}^2+rr'A'-BA}{BAr^2}\\
G^\theta_\theta&=&(4r''A^2B+2A''BAr-2B'r'A^2+2A'r'BA\nonumber\\
&&-A'B'Ar-A'{}^2Br)/(4rA^2B^2)\\
\label{etheta}
G^\phi_\phi&=&G^\theta_\theta
\end{eqnarray}
The prime denotes derivative with respect to $\lambda$. We now write down the well known Schwarzschild black hole metric in de Sitter space \cite{Gibbons:1977mu}
\begin{equation}
ds^2=-(1-\frac{2M}{r}-r^2)dt^2+\frac{dr^2}{1-\frac{2M}{r}-r^2}+r^2d\Omega .
\label{ds}
\end{equation}
where we set the Hubble parameter (and thus, essentially, the cosmological constant) to unity. By Birkhoff's theorem, this metric is the unique solution representing a black hole in de Sitter spacetime. We use this metric to describe two patches in the north and south hemisphere as shown in the Fig.~\ref{wormhole3} A. We locate the north pole at the origin, $r=0$.

Since the north and south hemispheres are copies of Eq.~(\ref{ds}), we should find the value of the coordinate $r$ that corresponds to the equator. We can now use the physical requirement which states that on the south hemisphere a test particle is attracted to the south pole, while on the north hemisphere a test particle is attracted to the north pole. This implies that there is no net force at the equator, and consequently, the gravitational acceleration must vanish at that location. Therefore, the required condition reads
\begin{equation}
\partial_r g_{tt}=0 .
\end{equation}
From Eq.~(\ref{ds}) we find the location of the equator at $r_0=M^{1/3}$. This is where we connect the north and south hemispheres.

The only remaining task is to find the explicit form of $r(\lambda)$ across the whole spacetime.
To find a match between the black hole geometry and the background de Sitter spacetime, we parameterize the
transformation as
\begin{equation}
(r-2M-r^3)=a^2\sin^2 \lambda .
\label{gtt}
\end{equation}
We choose this form because we need a periodic function to obtain a continuous metric connecting two identical patches, so that it reduces to the Einstein-Rosen coordinate transformation for the vanishing cosmological constant nearby $\lambda = 0$ or $\lambda = \pi$. If the transformation is not continuous at the equator, we need to introduce some extra matter contribution there.

Since the north and south hemispheres are connected, the location of the equator is at the maximum of $a^2\sin^2\lambda$, which is at $\lambda=\pi/2$.
From Eq.~(\ref{gtt}) and the radial location of the equator, $r_0=M^{1/3}$, we find the value of $a^2$ as
\begin{equation}
a^2=M^{1/3}-3M.
\end{equation}
Since $a^2\sin^2\lambda$ monotonically increases from $\lambda=0$ to $\lambda =\pi/2$, the left hand side of Eq.~(\ref{gtt}) must be a monotonically increasing function of $r$ from the black hole horizon to $r_0$. Therefore $M$ must satisfy
\begin{equation}
\label{range1}
M<3^{-\frac{3}{2}} .
\end{equation}
This condition puts a restriction on the black hole mass that gives a satisfactory solution. Since we set the Hubble parameter to one, this black hole mass is given in units of the inverse Hubble parameter, which in turn depends on the cosmological constant of
de Sitter space.
We can now write down the solution to Eq.~(\ref{gtt}) as
\begin{eqnarray}
r&=&\frac{\cos (\alpha)}{\sqrt{3}}-\sin (\alpha),\\
\alpha &=&\frac{\arctan\Big(\frac{\sqrt{3-81b^2}}{9b}\Big)}{3},\\
b&=&M+\frac{a^2}{2}\sin^2\lambda,
\end{eqnarray}
which gives the explicit  form of the coordinate $r(\lambda)$ in the metric (\ref{full-metric}).
The remaining metric elements in Eq.~(\ref{full-metric}) are
\begin{eqnarray}
A&=&1-\frac{2M}{r(\lambda)}-r^2(\lambda),\\\label{B}
B&=&\frac{1}{1-\frac{2M}{r(\lambda)}-r^2(\lambda)}\Big(\frac{dr(\lambda)}{d\lambda}\Big)^2 .
\end{eqnarray}

Note that the resulting metric looks like that of the Schwarzschild black hole in de Sitter space.
However, in the $r$ and $t$ coordinates one sees only the local geometry. By introducing
the $\lambda$  parameter we are nontrivially matching the two copies. The $\lambda$ parameter determines the physical metric across the whole spacetime, except at the equator, where $B=0$. To connect to the equator smoothly, we use the coordinate transformation 
\begin{equation}
r=r_0 - |\xi| ,
\end{equation}
which gives the following metric
\begin{equation}
ds^2=-(1-\frac{2M}{r}-r^2)dt^2+\frac{d\xi^2}{1-\frac{2M}{r}-r^2}+r^2d\Omega .
\end{equation}
In these coordinates the equator is located at $\xi=0$, while two black hole horizons are at $\xi=\pm (r_0-r_h)$, where $r_h$ denotes the radius of the black hole horizon. Since $\partial_\xi^2 r= -2\delta(\xi)$, a delta function appears in the Einstein tensor
\begin{eqnarray}
G^t_t\vert_{\xi=0}&=&\frac{-4\delta(\xi)}{rB}\\
G^\lambda_\lambda \vert_{\xi=0}&=&0\\
G^\theta_\theta\vert_{\xi=0}&=&G^\theta_\theta\vert_{\xi=0}=\frac{-2\delta(\xi)}{rB} .
\end{eqnarray}
We have kept only the terms with $\delta(\xi)$, since only the delta function remains after integration. The non-zero elements of the Einstein tensor indicate that there is a positive energy density and tension at the equator. The geometry of this matter distribution is described by a shell located at $\xi=0$, or equivalently $r=r_0$. Thus, to smoothly cover the whole spacetime, we need to use the $\lambda$ parametrization everywhere except around the equator, where we need to switch to the $\xi$ parameter.  

The null energy condition and the weak energy condition are easily checked by looking at
\begin{equation}
T_{\alpha\beta}x^\alpha x^\beta \ge 0
\end{equation}
where $x^\alpha$ is a null or timelike vector. In general relativity $G^\alpha_\beta=T^\alpha_\beta$, implying $|T^t_t|=2|T_\theta^\theta|=2|T^\phi_\phi|$. These two conditions are always satisfied at the equator. At the same time, since $|T^t_t|>|T_\theta^\theta|=|T^\phi_\phi|$, the dominant energy condition is satisfied as well. The strong energy condition requires
\begin{equation}
\label{strong}
\bar{T}_{\alpha\beta}x^\alpha x^\beta \ge 0
\end{equation}
where, $\bar{T}_{\alpha\beta}=(T_{\alpha\beta} -\frac{1}{2} tr(T) g_{\alpha\beta})$. The non-zero elements of $\bar{T}_{\alpha\beta}$ are
\begin{eqnarray}
\bar{T}^t_t\vert_{\xi=0}&=&\frac{-8\delta(\xi)}{rB}\\
\bar{T}^\lambda_\lambda \vert_{\xi=0}&=&\frac{4\delta(\xi)}{rB}\\
\bar{T}^\theta_\theta\vert_{\xi=0}&=&G^\theta_\theta\vert_{\xi=0}=\frac{2\delta(\xi)}{rB} 
\end{eqnarray}
It can be checked that equation (\ref{strong}) is satisfied. The strong energy condition is satisfied at the equator as well. Outside the mass shell, we have de Sitter vacuum, and the strong energy condition is violated, as usual.

 The total energy, $E_e$, at the equator can be found by integrating the $T_t^t=G^t_t$ near the equator:
\begin{equation}
E_e=-\int_{-\epsilon}^{\epsilon}G^t_t 4\pi \sqrt{B} r_0^2d\xi=16\pi M^{1/3}\sqrt{1-3M^{2/3}}.
\end{equation}
If $M>(-3 \times 2^7 \pi^2+\sqrt{3^2 2^{14}\pi^4 +2^8\pi^2})^{3/2}$, then $M>E_e$. This condition has an overlap with the equation (\ref{range1}). 
Therefore, 
it is possible to satisfy the condition 
$M>E_e$ or, it is possible to arrange for the black hole masses to be greater than the shell's energy.

This metric allows for a wormhole throat, located at $\lambda=0$ or $\xi=\pm(r_0-r_h)$, to connect the north and south pole directly, and not through the equator. The relevant geometry is shown in  Fig.~\ref{wormhole3} B. The global topology of the resulting spacetime is that of a torus. 
Since this is a static universe, any signal can causally propagate from one to the other black hole across the equator. However, the wormhole is not traversable because of the existence of the horizon.
Note that the solution connects two black holes in the same universe. Of course one can connect the north pole to any other black hole located anywhere in the other universe. This will make the spacetime structure more complicated.

We note that the following reference \cite{McInnes:2003xm} has considered some subtleties related to the Schwarzschild-de Sitter solution, however, without addressing the possibility of a wormhole solution. Our solution has some superficial similarities with the classic paper \cite{Gibbons:1977mu}, where a maximal extension of the Schwarzschild de Sitter solution was briefly discussed (see also \cite{ferrari}).
However, the crucial difference is that our construct does not have a cosmological horizon between the black holes, while this classic paper uses the full patch of the static de Sitter spacetime and connects different patches either at the black hole horizon or at the cosmological horizon. No useful information can travel through the cosmological or black hole horizons. Therefore even though the patches might be connected in \cite{Gibbons:1977mu}, observers located in different patches cannot exchange any useful information with other patches. Our solution does not use the whole static de Sitter spacetime. It includes only the spacetime from the black hole horizon to that place where the gravitational acceleration is $0$. Since the spacetime metric is not divergent, the signal can propagate, in principle, from one patch to the other one without any problems. If one can avoid the horizon, then a traversable wormhole may be possible \cite{Morris:1988cz}. In our case, the wormhole throat from one black hole to the other black hole is not traversable, because there is a usual Schwarzschild de Sitter black hole horizon. Note that there exist traversable wormhole solutions in the available literature \cite{Roman:1992xj,Ayon-Beato:2015eca,Canfora:2017gno}. A traversable wormhole may be constructed by replacing the metric nearby the black hole with a traversable wormhole solution. Note that our goal was to construct a wormhole solution such that outside of which any two points can communicate with each other. In other words,
the two mouths of our wormhole solution open into the same universe.
After this clarification, we collect some comments about the physical relevance of this solution.

1) First, it would be interesting to understand this solution from the
Euclidean point of view. Of course, the
Lorentzian solution is more physical, and more general, but the
corresponding Euclidean solution should be understood as well, especially in
the context of Euclidean quantum gravity.
2) Next, it would be natural to understand the doubly Wick rotated solution
in the AdS-context. The Lorentzian de Sitter and the Euclidean Anti-de Sitter have the same isometries, and thus it should be possible to
relate the Lorentzian de Sitter solution to the Euclidean Anti-de Sitter solution. As noted in \cite{Balasubramanian:2002zh} (see also \cite{Balasubramanian:2001nb}),
there exists a natural non-local map between these two spaces,
that can be used in this context.
3) This in turn leads to the issue of the possible holographic meaning
of the new wormhole solution. One natural guess is that this solution
represents an entangled state between two conformal field theory duals
living on the infinite past and future ``boundaries'', once again discussed
in the above paper \cite{Balasubramanian:2002zh}. In particular, it is natural to conjecture that this is a BCS-like
entangled state well known
from the BCS theory of superconductivity (and also mentioned in the
same paper \cite{Balasubramanian:2002zh}). Perhaps the 3d example is the simplest here, because it involves two entangled 2d Euclidean CFT duals.
The 3d example should be also understood from the point of view of the Chern-Simons formulation of 3d gravity.
4) One can also envision that the so-called elliptical de Sitter (going back to Schr\"{o}dinger's
classic book \cite{schrodinger}, in which the
antipodal points are identified, and the space is not time-orientable) should be important for the full understanding of the wormhole solution. The elliptical de Sitter was pursued in
\cite{Parikh:2002py}
in the early days of dS/CFT correspondence.
5) Also, in the context of Euclidean and Lorentzian quantum gravity, it would be valuable to understand multi-wormhole solutions. In the dilute wormhole gas approximation, such multi-wormhole solutions should be independent and
weakly interacting. In general, of course, the dense wormhole gas is not
a simple superposition of individual wormhole solutions.
6) Finally, any discussion of wormholes always involves the issues
of stability and causality, which are also subtle in the holographic context, especially in de Sitter space \cite{Balasubramanian:2001nb, Balasubramanian:2002zh}. In particular, the dual Euclidean CFT had unusual conjugacy relations for its Virasoro generators. Thus one could easily imagine that the stability of the wormhole bridge is related to such unusual properties of the
holographic dual.

Next, we collect some comments related to the cosmological constant problem and the relevance of topology change and wormhole configurations  (including, not only the new wormhole discussed in this letter, but also the standard maximally extended Schwarzschild de Sitter solution) in quantum gravity/string theory. 
Given such wormhole solutions in de Sitter space, it is tempting to think about the old Baum-Hawking-Coleman mechanism \cite{Baum:1984mc}, \cite{Hawking:1984hk}, \cite{Coleman:1988tj} for resolving the cosmological
constant problem from a new point of view, inspired by the recent discussion about wormholes and entanglement, or ER=EPR \cite{Maldacena:2013xja}, \cite{holland}.
We recall that, in particular, Coleman's version of the Baum-Hawking-Coleman mechanism \cite{Coleman:1988tj}, which asserts that topology change (in Euclidean quantum gravity)
may imply a probabilistic distribution for coupling constants in the relevant effective field theory (based on Euclidean quantum gravity coupled to matter)
and, in particular, that the relevant probability distribution (in the dilute wormhole approximation) for the cosmological constant is peaked around $0+$.
Nevertheless, there are many problems with this proposal:

A) Euclidean quantum gravity has many conceptual issues, and thus it might not be entirely trustworthy (even though the AdS/CFT correspondence has clarified some of its aspects), and the Lorentzian analysis gives complex phases, so the Baum-Hawking-Coleman measure for the cosmological constant is not peaked around any real value (as shown by Joe Polchinski in \cite{Polchinski:1989ae}). B) The proposal suffers from the menace of the giant wormholes (as pointed out by Kaplunovsky (unpublished) and Fischler and Susskind in \cite{Fischler:1988ia}). C) The proper analysis (claimed to have been done by Fischler, Klebanov, Polchinski and Susskind \cite{Fischler:1989ka}),
asserts that Coleman's proposal should be re-interpreted from the point of view of infrared divergences in quantum gravity (which connects to the current discussion about soft modes and deep infrared physics \cite{Strominger:2017zoo}).
D) There was a philosophical dissatisfaction that this proposal in some sense ignores short distance physics, even though it does
emphasize the role of non-locality in the cosmological constant problem.
(A few nice and balanced reviews on this subject are presented in \cite{Klebanov:1988eh} and \cite{Preskill:1988na}.
 For a discussion of the Baum-Hawking-Coleman measure and holography, see \cite{Horava:2000tb}.)

However, if we follow the idea of connecting holography and the Baum-Hawking-Coleman proposal \cite{Horava:2000tb} and the recent suggestion that wormholes (a la ER bridges) can be understood as being ``dual'' (or equivalent, in some sense)
to entanglement (a la EPR), the so called ``ER=EPR'' \cite{Maldacena:2013xja}, \cite{holland}, we could rephrase the Coleman proposal from the point of view of entanglement of
degrees of freedom at short distance and long distance (in a Lorentzian picture of quantum gravity).
Note that in this case the wormholes cannot be traversed.
In this way, one would end up with non-locality (the good feature) and perhaps evade the problem of giant wormholes. This could provide a proper Lorentzian proposal (in the Euclidean picture, the entanglement does
not really make sense, because we cannot define spacelike separated regions), and finally, we might connect to the recent discussion
of the relevance of soft modes in the infrared \cite{Strominger:2017zoo} in the context of quantum gravity and the cosmological constant problem.
Both the high energy and low energy modes should be essential from this new point of view. Thus, according to this new scenario, the (maximal) entanglement of high energy and low energy modes (and, thus, the maximal entanglement of the two holographically dual Euclidean CFTs) should be crucial for understanding why the universe is large.

In this way one would be turning the Baum-Hawking-Coleman proposal upside down in order to explore the good feature of non-locality via entanglement, while (hopefully)
avoiding the bad features of Euclidean quantum gravity and Euclidean effective field theory with topology change.
(We should also note the recent criticism of Euclidean quantum gravity and the need for a fundamental Lorentzian description in \cite{Feldbrugge:2017kzv}.)
There are also connections here with the recent research on quantum gravity/string theory \cite{Freidel:2013zga, Freidel:2014qna, Freidel:2015pka, Freidel:2015uug, Freidel:2016pls, Freidel:2017xsi, Freidel:2017wst, Freidel:2017nhg} with intrinsic non-locality, and other approaches to the cosmological constant problem, such as the sequester mechanism \cite{Kaloper:2014dqa}, which invoke the Coleman mechanism, at least, in spirit.  (For other connections between wormholes and cosmology, see, \cite{Chen:2016ask}.)

In conclusion, we have presented here a new  wormhole solution connecting two causally connected points of the same universe separated by finite distance.
This solution was constructed by placing two black holes at the antipodes
of the closed de Sitter space with a matter shell between them. By utilizing the
gravitational action of the matter shell and cosmological constant
which counteracts
attractive gravity between the black holes the whole configuration can be made static. 
The obtained spacetime does not have a cosmological horizon between the black holes which makes it substantially different from the maximal extension of the Schwarzschild de Sitter solution. The metric can be smooth at the equator, but some matter distribution with positive energy density must be placed there.
Motivated by this solution, we have then outlined its physical relevance in the
context of the relation between wormhole configurations and quantum entanglement which should be important for the Baum-Hawking-Coleman proposal, albeit from a new and more general viewpoint. We plan to explore these implications elsewhere.

{\it Acknowledgments:} We thank Alexander Vilenkin for discussions, and, especially, we thank Don Marolf for illuminating
emails. D.C Dai was supported by the National Science Foundation of China (Grant No. 11433001 and 11775140), National Basic Research Program of China (973 Program 2015CB857001) and  the Program of Shanghai Academic/Technology Research Leader under Grant No. 16XD1401600. D. M. is supported by the Julian Schwinger Foundation. D.M. would like to thank Laurent Freidel and Rob Leigh for numerous insightful discussions over many years on various ramifications of
quantum gravity. D.S. was partially supported by the US National Science Foundation, under Grant No. PHY-1820738.




\begin{thebibliography}{99}

\bibitem{ER} A. Einstein and N. Rosen, Phys. Rev. {\bf 48}, 73 (1935).

\bibitem{geons} J. A. Wheeler, 
Phys.\ Rev.\ {\bf 97}, 511 (1955).

\bibitem{wheeler} J. A. Wheeler, {\it Geometrodynamics}, Academic, New York, 1962.

\bibitem{Baum:1984mc}
  E.~Baum,
  Phys.\ Lett.\  {\bf 133B}, 185 (1983).

\bibitem{Hawking:1984hk}
  S.~W.~Hawking,
  Phys.\ Lett.\  {\bf 134B}, 403 (1984).

\bibitem{Coleman:1988tj}
  S.~R.~Coleman,
  Nucl.\ Phys.\ B {\bf 310}, 643 (1988).


\bibitem{gibbons}
G. W. Gibbons and S. W. Hawking (editors), {\it Euclidean Quantum Gravity}, World Scientific, 1993.

\bibitem{Morris:1988cz}
  M.~S.~Morris and K.~S.~Thorne,
  Am.\ J.\ Phys.\  {\bf 56}, 395 (1988).

\bibitem{Morris:1988tu}
  M.~S.~Morris, K.~S.~Thorne and U.~Yurtsever,
  Phys.\ Rev.\ Lett.\  {\bf 61}, 1446 (1988).


\bibitem{visser} M. Visser, {\it Lorentzian Wormholes: From Einstein to Hawking}, AIP Press, New York, 1995


\bibitem{Maldacena:2013xja}
  J.~Maldacena and L.~Susskind,
  Fortsch.\ Phys.\  {\bf 61}, 781 (2013)
  [arXiv:1306.0533 [hep-th]].


\bibitem{holland}
In P. R. Holland's book,  {\it Quantum Theory of Motion}, (Cambridge 1995), a connection between $ER$ and $EPR$ has been
suggested in the context of the de-Broglie-Bohm interpretation of quantum theory.


\bibitem{Hull:1998vg}
  C.~M.~Hull,
  JHEP {\bf 9807}, 021 (1998)
  [hep-th/9806146].


\bibitem{Strominger:2001pn}
  A.~Strominger,
  JHEP {\bf 0110}, 034 (2001)
  [hep-th/0106113].

\bibitem{Balasubramanian:2001rb}
  V.~Balasubramanian, P.~Horava and D.~Minic,
  JHEP {\bf 0105}, 043 (2001)
  [hep-th/0103171].

\bibitem{Witten:2001kn}
  E.~Witten,
Proceedings of Strings 2001, Clay Math. Proc. {\bf 1} (2002),
  hep-th/0106109.


\bibitem{Balasubramanian:2001nb}
  V.~Balasubramanian, J.~de Boer and D.~Minic,
  Phys.\ Rev.\ D {\bf 65}, 123508 (2002)
  [hep-th/0110108].


\bibitem{Balasubramanian:2002zh}
  V.~Balasubramanian, J.~de Boer and D.~Minic,
  Class.\ Quant.\ Grav.\  {\bf 19}, 5655 (2002)
  [Annals Phys.\  {\bf 303}, 59 (2003)]
  [hep-th/0207245].


\bibitem{Maldacena:1997re}
  J.~M.~Maldacena,
  Int.\ J.\ Theor.\ Phys.\  {\bf 38}, 1113 (1999)
  [Adv.\ Theor.\ Math.\ Phys.\  {\bf 2}, 231 (1998)]
  [hep-th/9711200].


\bibitem{Gubser:1998bc}
  S.~S.~Gubser, I.~R.~Klebanov and A.~M.~Polyakov,
  Phys.\ Lett.\ B {\bf 428}, 105 (1998)
  [hep-th/9802109].

\bibitem{Witten:1998qj}
  E.~Witten,
  Adv.\ Theor.\ Math.\ Phys.\  {\bf 2}, 253 (1998)
  [hep-th/9802150].




\bibitem{Gibbons:1977mu}
  G.~W.~Gibbons and S.~W.~Hawking,
  Phys.\ Rev.\ D {\bf 15}, 2738 (1977).


\bibitem{McInnes:2003xm}
  B.~McInnes,
  JHEP {\bf 0309}, 009 (2003)
  [hep-th/0308022].

\bibitem{ferrari}
S. L. Bazanski and V. Ferrari,  Il Nuovo Cimento, {\bf 91 B} 126 (1986).


\bibitem{Roman:1992xj}
  T.~A.~Roman,
  Phys.\ Rev.\ D {\bf 47}, 1370 (1993)
  [gr-qc/9211012].



\bibitem{Ayon-Beato:2015eca}
  E.~Ayon-Beato, F.~Canfora and J.~Zanelli,
  Phys.\ Lett.\ B {\bf 752}, 201 (2016)
  [arXiv:1509.02659 [gr-qc]].

\bibitem{Canfora:2017gno}
  F.~Canfora, N.~Dimakis and A.~Paliathanasis,
  Phys.\ Rev.\ D {\bf 96}, no. 2, 025021 (2017)
  [arXiv:1707.02270 [hep-th]].



\bibitem{schrodinger} E. Schr\"{o}dinger, {\it Expanding Universe},  Cambridge, 1956 (2011 reprint).

\bibitem{Parikh:2002py}
  M.~K.~Parikh, I.~Savonije and E.~P.~Verlinde,
  Phys.\ Rev.\ D {\bf 67}, 064005 (2003)
  [hep-th/0209120].




\bibitem{Polchinski:1989ae}
  J.~Polchinski,
  Nucl.\ Phys.\ B {\bf 325}, 619 (1989).

\bibitem{Fischler:1988ia}
  W.~Fischler and L.~Susskind,
  Phys.\ Lett.\ B {\bf 217}, 48 (1989).

 \bibitem{Fischler:1989ka}
W.~Fischler, I.~R.~Klebanov, J.~Polchinski and L.~Susskind,
  Nucl.\ Phys.\ B {\bf 327}, 157 (1989).


\bibitem{Strominger:2017zoo}
  A.~Strominger,
  {\it Lectures on the Infrared Structure of Gravity and Gauge Theory}, Princeton, 2018.
  arXiv:1703.05448 [hep-th].





\bibitem{Klebanov:1988eh}
  I.~R.~Klebanov, L.~Susskind and T.~Banks,
  Nucl.\ Phys.\ B {\bf 317}, 665 (1989).

\bibitem{Preskill:1988na}
  J.~Preskill,
  Nucl.\ Phys.\ B {\bf 323}, 141 (1989).


\bibitem{Horava:2000tb}
  P.~Horava and D.~Minic,
  Phys.\ Rev.\ Lett.\  {\bf 85}, 1610 (2000)
  [hep-th/0001145].


\bibitem{Feldbrugge:2017kzv}
  J.~Feldbrugge, J.~L.~Lehners and N.~Turok,
  Phys.\ Rev.\ D {\bf 95}, no. 10, 103508 (2017)
  [arXiv:1703.02076 [hep-th]].




\bibitem{Freidel:2013zga}
L.~Freidel, R.~G.~Leigh and D.~Minic,
  Phys.\ Lett.\ B {\bf 730}, 302 (2014)
  [arXiv:1307.7080 [hep-th]].

\bibitem{Freidel:2014qna}
L.~Freidel, R.~G.~Leigh and D.~Minic,
  Int.\ J.\ Mod.\ Phys.\ D {\bf 23}, no. 12, 1442006 (2014)
  [arXiv:1405.3949 [hep-th]].

\bibitem{Freidel:2015pka}
  L.~Freidel, R.~G.~Leigh and D.~Minic,
  JHEP {\bf 1506}, 006 (2015)
  [arXiv:1502.08005 [hep-th]].

\bibitem{Freidel:2015uug}
  L.~Freidel, R.~G.~Leigh and D.~Minic,
  Int.\ J.\ Mod.\ Phys.\ D {\bf 24}, no. 12, 1544028 (2015).


\bibitem{Freidel:2016pls}
  L.~Freidel, R.~G.~Leigh and D.~Minic,
  Phys.\ Rev.\ D {\bf 94}, no. 10, 104052 (2016)
  [arXiv:1606.01829 [hep-th]].

\bibitem{Freidel:2017xsi}
  L.~Freidel, R.~G.~Leigh and D.~Minic,
  J.\ Phys.\ Conf.\ Ser.\  {\bf 804}, no. 1, 012032 (2017).


\bibitem{Freidel:2017wst}
  L.~Freidel, R.~G.~Leigh and D.~Minic,
  JHEP {\bf 1709}, 060 (2017)
  [arXiv:1706.03305 [hep-th]].

\bibitem{Freidel:2017nhg}
  L.~Freidel, R.~G.~Leigh and D.~Minic,
  Phys.\ Rev.\ D {\bf 96}, no. 6, 066003 (2017)
  [arXiv:1707.00312 [hep-th]].


\bibitem{Kaloper:2014dqa}
  N.~Kaloper and A.~Padilla,
  Phys.\ Rev.\ D {\bf 90}, no. 8, 084023 (2014)
  Addendum: [Phys.\ Rev.\ D {\bf 90}, no. 10, 109901 (2014)]
  [arXiv:1406.0711 [hep-th]].


\bibitem{Chen:2016ask}
  P.~Chen, Y.~C.~Hu and D.~h.~Yeom,
  JCAP {\bf 1707}, no. 07, 001 (2017)
  [arXiv:1611.08468 [gr-qc]].


\end{thebibliography}
\end{document}